\documentclass[twocolumn, a4paper,14pt, showkeys, showpacs]{revtex4}
\usepackage{amsmath}
\usepackage{amssymb}
\usepackage{amsmath,amsthm,amssymb,amscd}
\usepackage{latexsym}
\usepackage{indentfirst}
\usepackage{subfigure}
\usepackage{graphicx}
\usepackage{extarrows}
\usepackage{bm}
\usepackage{multirow}
\usepackage[colorlinks,linkcolor=blue,hyperindex,dvipdfm]{hyperref}

\date{\today}

\begin{document}

\title{  \bf Mean-field analysis of two-species TASEP with attachment and detachment
}

\author{Minghua Song and Yunxin Zhang}
\email[Email: ]{xyz@fudan.edu.cn}
\affiliation{Laboratory of Mathematics for Nonlinear Science, Shanghai Key Laboratory for Contemporary Applied Mathematics, Centre for Computational Systems Biology, School of Mathematical Sciences, Fudan University, Shanghai 200433, China. }

\begin{abstract}
In cells, most of cargos are transported by motor proteins along microtubule. Biophysically, unidirectional motion of large number of motor proteins along a single track can be described by totally asymmetric simple exclusion process (TASEP). From which many meaningful properties, such as the appearance of domain wall (defined as the borderline of high density and low density of motor protein along motion track) and boundary layers, can be obtained. However, it is biologically obvious that a single track may be occupied by different motor species. So previous studies based on TASEP of one particle species are not reasonable enough to find more detailed properties of the motion of motors along a single track. To address this problem, TASEP with two particle species is discussed in this study. Theoretical methods to get densities of each particle species are provided. Using these methods, phase transition related properties of particle densities are obtained. Our analysis show that domain wall and boundary layer of single species densities always appear simultaneously with those of the total particle density. The height of domain wall of total particle density is equal to the summation of those of single species. Phase diagrams for typical model parameters are also presented. The methods presented in this study can be generalized to analyze TASEP with more particle species.

\end{abstract}



\maketitle

Many driven diffusion systems have been developed to model intracellular motility \cite{Bressloff2013,Chowdhury2013}. Among which, one-dimensional totally asymmetric simple exclusion process (TASEP) is usually employed to describe the unidirectional motion of large number of motor proteins along microtubule, the transcription process of RNA polymerases along DNA, and the translation process of ribosomes along messenger RNA transcript \cite{Nagar2011,Leduc2012,Racle2013}. In TASEP, motion track of particles is simplified to be a one-dimensional lattice with length $N+1$, particles  enter the track at first/initiation site $0$ provided site $0$ is not occupied, and leave track from last/termination site $N$. If lattice site $i+1$ is not occupied, the particle at lattice site $i$ will hop forward to site $i+1$ with given rate. Generally, particles may also detach into environment and end their unidirectional motion from bulk lattice sites, and new particles may attach to any one of unoccupied bulk sites. In last decades, TASEP has been studied extensively, especially for phase transition related properties, i.e., the appearance of domain wall (DW) and boundary layer (BL), which are usually driven by boundary conditions \cite{Derrida1992,Parmeggiani2003,Pronina2005,Nishinari2007,Blythe2007,Zhang2012,Matsui2015}.
However, in most of previous studies, the self-propelled particles are usually assumed to be from the same species. But in cells, one protofilament of microtubule may be occupied by motor proteins from different species \cite{Vale2003}. So, to know more detailed properties about the motion of motor proteins in real cells, the usual TASEP should be generalized to include particles from different species \cite{Alcaraz1998,Arita1999,Ferrari2007,Arita2013}. 

The simplest generalization is two-species ASEP, which has been discussed in recent studies \cite{Evans1995,Mobilia2001,Schutz2003,Zeraati2013}. However, in almost all previous studies, no particle attachment/detachment is allowed to/from bulk sites of the track, i.e. the site $i$ for $1\le i\le N-1$. They usually assumed that one species enters track at site $0$ and leaves from site $N$, while the other species enters at site $N$ and leaves from site $0$. Meanwhile, it is also assumed that two-species pair $(P_1)_i(P_2)_{i+1}$ can change to $(P_2)_i(P_1)_{i+1}$, i.e. forward hopping of one species is not blocked by the other species. Here $(P_k)_i$ means there is a particle $P_k$ at site $i$. For convenience, the two particle species are denoted by $P_1$ and $P_2$ respectively.

The TASEP discussed in this study also includes two particle species. But both of them enter track at initiation site $0$ and leave from termination site $N$, i.e. they not only travel along the same track, but also move to the same direction. Previous studies about one species TASEP have shown that nontrivial attachment/detachment is one of the key driven factors to the appearance of DW in particle density along track \cite{Derrida1993,Schutz1993,Parmeggiani2003,Nishinari2005,Leduc2012,Zhang2012}. So, in this study, both of the two species are allowed to attach to (and detach from) bulk sites of the track. The same as in one species cases, we also called this process \lq\lq TASEP-LK'' process \cite{Parmeggiani2003}. 

Let $n_i$ and $m_i$ be occupation numbers of species $P_1$ and $P_2$ at site $i$, respectively. Specifically $n_i=1$ means site $i$ is occupied by a particle $P_1$, while $n_i=0$ means site $i$ is not occupied by particle $P_1$. Because of the hard-core exclusion, $n_i+m_i=0$ or 1. For $1\le i\le N-1$, the time evolution of $n_i$ and $m_i$ are governed by following equations
\begin{eqnarray}
\label{EqBulk:1}
    \nonumber dn_i/dt &=& n_{i-1}(1-n_i-m_i)-n_i(1-n_{i+1}-m_{i+1}) \\
                    && +\omega_{1,A}(1-n_i-m_i)-\omega_{1,D}n_i, \\
\label{EqBulk:2}
    \nonumber dm_i/dt &=& qm_{i-1}(1-n_i-m_i)-qm_i(1-n_{i+1}-m_{i+1}) \\
                    && +\omega_{2,A}(1-n_i-m_i)-\omega_{2,D}m_i.
\end{eqnarray}
Where, for convenience, forward hopping rate of species $P_1$ is normalized to be unit, and forward hopping rate of species $P_2$ is assumed to be $q\le1$. The attachment rate of species $P_k$ to any unoccupied bulk site $i$ is denoted by $\omega_{k,A}$, and the detachment rate of species $P_k$ from bulk sites is denoted by $\omega_{k,D}$. At initiation site $i=0$ (left boundary),
\begin{eqnarray}
\label{EqLeftBoundary:1}
  dn_0/dt &=& \alpha_1(1-n_0-m_0)-  n_0(1-n_{1}-m_{1}), \\
\label{EqLeftBoundary:2}
  dm_0/dt &=& \alpha_2(1-n_0-m_0)-qm_0(1-n_{1}-m_{1}).
\end{eqnarray}
Where $\alpha_k$ is the entry rate of particle $P_k$ from environment. While at termination site $i=N$ (right boundary),
\begin{eqnarray}
\label{EqRightBoundary:1}
  dn_N/dt &=&   n_{N-1}(1-n_N-m_N)-\beta_1n_N,\\
\label{EqRightBoundary:2}
  dm_N/dt &=& qm_{N-1}(1-n_N-m_N)-\beta_2m_N,
\end{eqnarray}
with $\beta_k$ the leaving rate of particle $P_k$ to environment.

In cells, hopping rate of particles is usually determined by their biochemical properties. For example, forward hopping of motor proteins, such as conventional kinesin, is mechanochemically coupled with ATP hydrolysis. Each forward mechanical step is tightly coupled with one ATP hydrolysis. Therefore, hopping rate is determined by the rate of ATP hydrolysis. Meanwhile, leaving rate of particles from motion track is also determined by their biochemical properties, or even the rate of ATP hydrolysis. Therefore it is biophysically reasonable to assume that the ratio of leaving rate of the two species $\beta_1:\beta_2$ is equal to ratio of their forward hopping rate $1:q$. Meanwhile, experiments found that, if the two heads of motor protein kinesin are both in ADP binding state, then it will soon detach from microtubule. This means that detachment rate of motor proteins are also determined by their biochemical properties and the rate of ATP hydrolysis. So, for convenience of theoretical analysis, this study also assumes that $\omega_{1,D}:\omega_{2,D}=1:q$. However, except their biochemical properties, entry rate $\alpha_k$ and attachment rate $\omega_{k,A}$ of species $P_k$ are also influenced by environmental conditions, especially their concentrations. So, the corresponding rate ratios may be different from the ratio $1:q$ of forward hopping rate.

Defining $\rho_1(i)=\langle n_i\rangle$ and $\rho_2(i)=\langle m_i\rangle$. By mean-field approximation, for large track length $N$ limit, equations for steady state values of densities $\rho_1$ and  $\rho_2$ can be obtained from Eqs. (\ref{EqBulk:1},\ref{EqBulk:2}),
\begin{eqnarray}
\label{EqBulk0}
    \partial_x J_k&=&q^{1-k}\Omega_{k,A}(1-\rho_1 -\rho_2 )-\Omega\rho_k, \quad k=1,2.
\end{eqnarray}
Where $J_k= \rho_k(1-\rho_1 -\rho_2)$, 
$\Omega_{k,A}=N\omega_{k,A}$, $\Omega=N\omega_{1,D}=N\omega_{2,D}/q=:N/\omega$, and $0<x<1$. Note, this study assumes that $\langle n_im_{i+1}\rangle=\rho_1(i)\rho_2(i+1)$, and in MFA analysis the track length is normalized to be 1. From Eq. (\ref{EqBulk0}), one can show that total particle density $\rho=\rho_1+\rho_2$ satisfies
\begin{equation}\label{EqBulk00:Sum}
  \partial_x J=K\Omega(1-\rho)-\Omega\rho,
\end{equation}
where $K=K_1+K_2=[\omega_{1,A}+\omega_{2,A}/q]/\omega$, and $J=\rho(1-\rho)$. From Eqs. (\ref{EqLeftBoundary:1}-\ref{EqRightBoundary:2}), one can show that at boundaries $x=0,1$, particle densities $\rho_1, \rho_2$ and $\rho$ satisfy $\rho_1(0)=\alpha_1,\rho_2(0)=\alpha_2/q$, and $\rho(1)=\rho_1(1)+\rho_2(1)=1-\beta$ with $\beta=\beta_1=\beta_2/q$.

\begin{figure}[!htp]
  \centering
  \includegraphics[width=8.6cm]{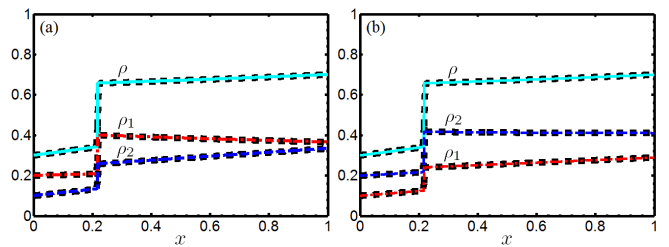}\\
  \caption{(Color online) Mean density profiles obtained from Eqs. (\ref{EqBulk:1}-\ref{EqRightBoundary:2})(black dotted line), and their mean-field approximations obtained from Eqs. (\ref{EqBulk0},\ref{EqBulk00:Sum})
  (cyan solid lines for $\rho$, red dashdot lines for $\rho_1$, and blue dashed line for $\rho_2$).
  Parameter values used in calculations are {\bf (a)} $\Omega_{1,A}=0.01$, $\Omega_{2,A}=0.14$, $\alpha_1=0.2$, $\alpha_2=0.09$, and {\bf (b)} $\Omega_{1,A}=0.1$, $\Omega_{2,A}=0.05$, $\alpha_1=0.1$, $\alpha_2=0.18$. Other parameter values are $N=10^3$, $q=0.9$,   $\Omega_{1,D}=\Omega_{2,D}/q=0.1$, $\beta_1=\beta_2/q=0.3$.
  Total particle density $\rho$ in both {\bf (a)} and {\bf (b)} is the same as the one in usual \lq\lq TASEP-LK" process with $\alpha=\beta=0.3$, and $\Omega_D=0.1$,  $\Omega_A=0.15$. Locations of domain wall for total density $\rho$ and single species densities $\rho_1, \rho_2$ are the same. Across domain wall location, current $J=\rho(1-\rho)$ and current $J_k=\rho_k(1-\rho)$ for $k=1,2$ are all conserved. }
  \label{Fig.Kisnot1_ex1}
\end{figure}

Eq. (\ref{EqBulk00:Sum}) implies that the governing equation for total density $\rho$ is the same as the one in one-species \lq\lq TASEP-LK" process \cite{Parmeggiani2004}. But with {\it effective} detachment rate $\Omega_D=\Omega$, {\it effective} attachment rate $\Omega_A=K\Omega$, initiation (entry) rate $\alpha=\alpha_1+\alpha_2/q$, and termination (leaving) rate $\beta=\beta_1=\beta_2/q$ (see Fig.\ref{Fig.Kisnot1_ex1}). So total density $\rho$ can be obtained by the same method as in one-species \lq\lq TASEP-LK" process. But the main difficulty for two-species cases is how to get single species densities $\rho_1$ and $\rho_2$. Actually, no reasonable boundary conditions at $x=1$, i.e., values of $\rho_1(1)$ and $\rho_2(1)$, can be derived from Eqs. (\ref{EqRightBoundary:1},\ref{EqRightBoundary:2}). Meanwhile, properties of densities $\rho_1$ and $\rho_2$ are different from the ones in one-species cases. For example, in one-species \lq\lq TASEP-LK" process, particle densities before and after DW location $x_w$ satisfies $\rho(x_w^-)+\rho(x_w^+)=1$. This is because that across location $x_w$, current $J=\rho(1-\rho)$ is conserved. But for two-species process, particle densities $\rho_1, \rho_2$ do not satisfy this relation. Instead,  conservation of current $J_k=\rho_k(1-\rho)$ gives that across DW location, density ratio $\rho_1/\rho_2$ is not changed (see Fig.\ref{Fig.Kisnot1_ex1}). The plots in Fig.\ref{Fig.Kisnot1_ex1} imply that DWs of density $\rho, \rho_1, \rho_2$ appear at the same location. Further numerical calculations show that their BLs also appear simultaneously, see Fig. S4 in \cite{supplemental}.

Along motion track, density currents $J, J_1, J_2$ always change continuously, even if corresponding densities $\rho, \rho_1, \rho_2$ are discontinuous. At any location $x$ of the track, $\rho(x^-)[1-\rho(x^-)]=J(x^-)=J(x^+)=\rho(x^+)[1-\rho(x^+)]$, and
$\rho_k(x^-)[1-\rho(x^-)]=J_k(x^-)=J_k(x^+)=\rho_k(x^+)[1-\rho(x^+)]$. Through simple analysis, we obtained $\rho_k(x^-)/\rho(x^-)=\rho_k(x^+)/\rho(x^+)$ or $\rho_1(x^-)/\rho_2(x^-)=\rho_1(x^+)/\rho_2(x^+)$.  This relation also holds at DW location $x_w$.

As we have mentioned before, total density $\rho$ can be obtained from Eq. (\ref{EqBulk00:Sum}) with boundary conditions $\rho(0)=\rho_1(0)+\rho_2(0)=\alpha_1+\alpha_2/q$ and $\rho(1)=1-\beta$. But, if DW exists between boundaries $x=0$ and $x=1$, then without boundary value $\rho_k(1)$, density $\rho_k$ cannot be directly determined by Eq. (\ref{EqBulk0}). Actually, with boundary value $\rho_k(0)$, only the value of density $\rho_k$ before DW location $x_w$ can be directly obtained by Eq. (\ref{EqBulk0}). One of the main aims of this study to find methods to get single species density $\rho_k$ along the whole track from Eq. (\ref{EqBulk0}), but with only boundary values at $x=0$.

Let $\Delta:=\rho(x_w^+)-\rho(x_w^-)$, i.e. the DW height of total density $\rho$. Then $\Delta_k:=\rho_k(x_w^+)-\rho_k(x_w^-)=
[\rho_k(x_w^+)/\rho(x_w^+)]\rho(x_w^+)-[\rho_k(x_w^-)/\rho(x_w^-)]\rho(x_w^-)
=[\rho_k(x_w^+)/\rho(x_w^+)]\Delta=[\rho_k(x_w^-)/\rho(x_w^-)]\Delta$. So, $\Delta=\Delta_1+\Delta_2$, and $\Delta>0$ iff $\Delta_k>0$. Therefore, DWs of total density $\rho$ and single species densities $\rho_1$ and $\rho_2$ always appear at the same location, which is consistent with the finding in numerical calculations (see Fig. \ref{Fig.Kisnot1_ex1}).

The value $\rho_k(x_w^+)$ can be obtained as follows, $\rho_k(x_w^+)=\rho_k(x_w^-)+\Delta_k=
\rho_k(x_w^-)+[\rho_k(x_w^-)/\rho(x_w^-)][\rho(x_w^+)-\rho(x_w^-)]=
\rho(x_w^+)\rho_k(x_w^-)/\rho(x_w^-)$. Using $\rho_k(x_w^+)$ as left boundary condition, the single species density $\rho_k$ after DW location $x_w$ can be obtained from  Eq. (\ref{EqBulk0}).
Note, the above method to get single species density $\rho_k$ is also applicable for the cases where there exists BL in density $\rho$. 
BLs of single species densities $\rho_1, \rho_2$ and total density $\rho$ are also appear simultaneously.

\begin{figure}[!htp]
  \centering
  \includegraphics[width=8.6cm]{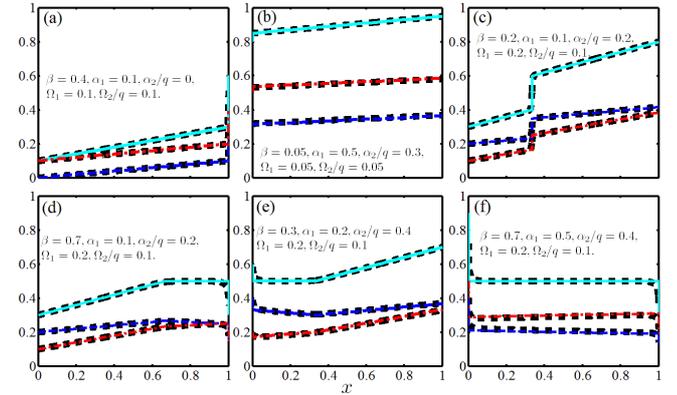}\\
  \caption{(Color online) Examples of density profiles for $K=1$. The same line types as in Fig. \ref{Fig.Kisnot1_ex1} are used.
  {\bf (a)} LD-BL phase with $\alpha<0.5$ and $\beta<0.5$;
  {\bf (b)} BL-HD phase with $\alpha>0.5$ and $\beta<0.5$;
  {\bf (c)} LD-DM-HD phase with $\alpha<0.5$ and $\beta<0.5$;
  {\bf (d)} LD-MC-BL phase with $\alpha<0.5$ and $\beta>0.5$;
  {\bf (e)} BL-MC-HD phase with $\alpha>0.5$ and $\beta<0.5$;
  {\bf (f)} BL-MC-BL phase with $\alpha>0.5$ and $\beta>0.5$.
  Where $\alpha=\alpha_1+\alpha_2/q$, and $\beta=\beta_1=\beta_2/q$. Unless explicitly presented, in all the following calculations $q=0.9$ is used.
  }
  \label{Fig.Kis1_2+3}
\end{figure}

If $K=1$,  Eq. (\ref{EqBulk00:Sum}) reduces to
\begin{eqnarray}\label{EqBulk00K=1}
  (\partial_x\rho-\Omega)(2\rho-1)&=&0,
\end{eqnarray}
which yields two solutions. The constant solution $\rho\equiv1/2$ coincides with the density $\rho_l=K/(K+1)$ given by Langmuir kinetics (LK), and corresponds to the maximal current (MC).
The other solution which matches left or right boundary condition is $\rho_\alpha(x)=\Omega x+\alpha$ or $\rho_\beta(x)=\Omega x+1-\beta-\Omega$.
Using these three solution candidates and based on the continuity of currents $J, J_1, J_2$, total density $\rho$ and single species densities $\rho_1, \rho_2$ can be obtained \cite{supplemental}. 
Typical examples of $\rho, \rho_1, \rho_2$ for $K=1$ are plotted in Fig. \ref{Fig.Kis1_2+3}, including the cases of low density (LD) phase ($\rho<1/2$), high density (HD) phase ($\rho>1/2$), the maximal current (MC) phase ($\rho=1/2$), as well as left and right BLs. Where thick dotted lines are obtained by numerical iterations of Eqs. (\ref{EqBulk:1}-\ref{EqRightBoundary:2}), and others are obtained from Eqs. (\ref{EqBulk0},\ref{EqBulk00:Sum}) using the method presented here \cite{supplemental}. 
Fig. \ref{Fig.Kis1_2+3} shows that for large initiation rate $\alpha=\alpha_1+\alpha_2>0.5$, left BL exists in both total density $\rho$ and single species densities $\rho_1$ and $\rho_2$.
Meanwhile, right BL may appear for either large or small values of termination rate $\beta$. Again, DW of single species density $\rho_k$ occurs at the same location as that of total density $\rho$. Which satisfies $\Delta_k=[\rho_k(x_w^-)/\rho(x_w^-)]\Delta$ and $\Delta=\Delta_1+\Delta_2$.

For special cases $K=1$, phases of single species densities $\rho_1$ and $\rho_2$ may not be the same as that of total density $\rho$. If $\rho$ is in LD phase, then both $\rho_1$ and $\rho_2$ will be in LD phase, see Fig. \ref{Fig.Kis1_2+3}{\bf (a)}. However, when $\rho$ is in HD phase, one of the single species densities, or even both of them, may still remain in LD phase, see Fig. \ref{Fig.Kis1_2+3}{\bf (b,e)}. For special cases $K=1$, if initiation rate $\alpha$ and termination rate $\beta$ satisfy \cite{supplemental}. 
\begin{equation}\label{EqMCcondition}
      \alpha >1/2-\Omega, \quad
      \beta  >1/2-\Omega,\quad \textrm{and }\quad
      \alpha+\beta>1-\Omega,
\end{equation}
then density $\rho$ may be in maximal current phase (MC) near right boundary $x=1$, in which $\rho\equiv1/2$ and current $J=J_1+J_2=1/4$.
Depends on rates $\alpha_k$ and $\Omega_{k,A}$, single species density $\rho_k$ may not be constant, and they may increase or decrease along motion track. This is different from the cases in which density $\rho$ is in LD or HD phase. For those cases both $\rho_1$ and $\rho_2$ increase with $x$, and with slops $\Omega_{1,A}$ and $\Omega_{2,A}/q$, respectively, see Eqs. (S7,S7) in \cite{supplemental}.
It can be shown that when $\rho\equiv1/2$, slopes of $\rho_1$ and $\rho_2$ have same absolute value, but opposite signs. If $\Omega_{1,A}\alpha_2-\Omega_{2,A}\alpha_1>0$ then $\rho_1$ has positive slope, i.e. increases along the motion track, while $\rho_2$ has negative slope. Both $\rho_1, \rho_2$ are also constants iff $\Omega_{1,A}\alpha_2-\Omega_{2,A}\alpha_1=0$, see 
Fig. S2 in \cite{supplemental}. 

Based on current continuity, and using expressions of $\rho_\alpha$ and $\rho_\beta$, which satisfy Eq. (\ref{EqBulk0}) with boundary conditions $\rho_\alpha(0)=\alpha$ and $\rho_\beta(1)=\beta$ respectively, we found that for $K=1$, DW appears iff $|\alpha-\beta| <\Omega$ and $\alpha+\beta+\Omega<1$ \cite{supplemental}. 
The DW lies at $x_w=(\Omega+\beta-\alpha)/(2\Omega)$ with height $\Delta=1-\alpha-\beta-\Omega$. The DW height of density $\rho_k$ can then be obtained by $\Delta_k=[\rho_k(x_w^-)/\rho(x_w^-)]\Delta$, see Eqs. (S4,S5,S7,S8) in \cite{supplemental}.

\begin{figure}[!htp]
  \centering
  \includegraphics[width=8.6cm]{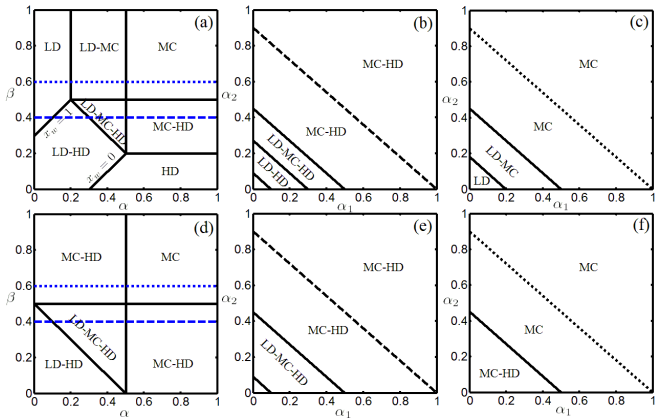}\\
  \caption{(Color online) Phase diagrams of total density $\rho=\rho_1+\rho_2$ obtained from stationary mean-field Eq.(\ref{EqBulk00:Sum}) with $K=1$, $\Omega=0.3$ in {\bf (a,b,c)}, and $\Omega=0.5$ in {\bf (d,e,f)}. {\bf (b,c)} and {\bf (e,f)} are phase diagrams in $(\alpha_1,\alpha_2)$ plane, which correspond to $\beta=0.4$ and $\beta=0.6$, respectively (see the horizontal dotted and dashed lines in {\bf (a,d)}). 
  }
  \label{Fig.Kis1Phase}
\end{figure}

Using similar method as in \cite{Parmeggiani2004}, typical examples of phase diagram of density $\rho$, for special cases $K=1$, are given in Fig. \ref{Fig.Kis1Phase}. Generally, density $\rho$ may be in any one of the seven possible phases, which include all combinations of LD, MC, and HD. The lines in Figs. \ref{Fig.Kis1Phase}{\bf (b,c,e,f)} are obtained by $\alpha_2=q(\alpha-\alpha_1)$, since density $\rho$ depends only on the summation $\alpha=\alpha_1+\alpha_2/q$.

As mentioned in \cite {Parmeggiani2003}, due to particle-hole symmetry we only need to discuss the \lq\lq TASEP-LK" process for $K\geq1$. For $K>1$, the idea used in the special cases $K=1$ can also be employed to get single special density $\rho_k$ \cite{supplemental}. Roughly speaking, total density $\rho$, as well as its locations of DW and BL, can be obtained from Eq. (\ref{EqBulk00:Sum}) with boundary conditions $\rho(0)=\alpha=\alpha_1+\alpha_2/q$ and $\rho(1)=1-\beta=1-\beta_1=1-\beta_2/q$. Then single species density $\rho_k$ in interval $[0, x_w)$ can be obtain from Eq. (\ref{EqBulk0}) with left boundary condition $\alpha_1$ or $\alpha_2/q$. Finally, density $\rho_k$ in interval $(x_w,1]$ can be obtained from Eq. (\ref{EqBulk0}) with left boundary condition $\rho_k(x_w^+)$, which is given by $\rho_k(x_w^+)=\rho_k(x_w^-)+\Delta_k$. The main difference from that in special cases $K=1$ is that Lambert function \cite{Lambert} should be employed to help to get densities $\rho$ and $\rho_k$ \cite{supplemental}.

\begin{figure}[!htp]
  \centering
  \includegraphics[width=8.6cm]{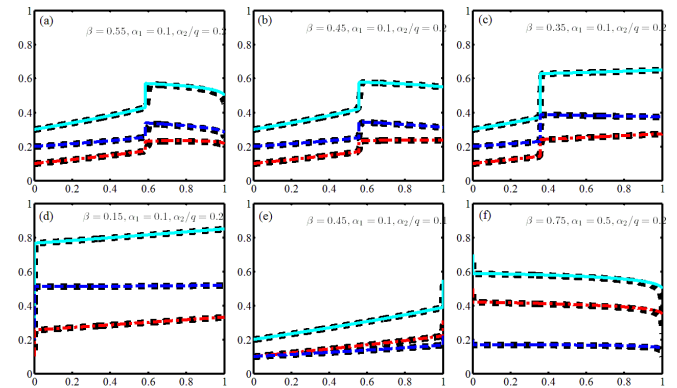}\\
  \caption{(Color online) Typical examples of density $\rho$ for $\Omega=0.1$ and $K=1.5$, i.e. $\rho_l=K/(K+1)=0.6$.
  {\bf (a)} LD-DW-HD phase, with $\alpha<0.5$ and $\beta>0.5$;
  {\bf (b)} LD-DW-HD phase, with $\alpha<0.5$ and $1-\rho_l<\beta<0.5$;
  {\bf (c)} LD-DW-HD phase, with $\alpha<0.5$ and $0<\beta<1-\rho_l$;
  {\bf (d)} HD phase, with $\alpha<0.5$ and $\beta<1-\rho_l$;
  {\bf (e)} LD phase, with $\alpha<0.5$ and $1-\rho_l<\beta<0.5$;
  {\bf (f)} \lq\lq Meissner" phase, with $\alpha\geq 0.5$ and $\beta\geq 0.5$. Line types are the same as in Fig. \ref{Fig.Kisnot1_ex1}.
  }
  \label{Fig.Analysis_vs_Simulation_K_is_not_1_ex}
\end{figure}

For $K>1$, properties of total density $\rho$ are similar as the ones of the one-species \lq\lq TASEP-LK" process \cite{Parmeggiani2003,Zhang2012}, i.e. there may exist left or right boundary layer, domain wall, or \lq\lq Meissner'' (M) phase, see Fig. \ref{Fig.Analysis_vs_Simulation_K_is_not_1_ex}. Here \lq\lq Meissner'' phase means that density $\rho$ satisfies $1/2<\rho<\rho_l=K/(K+1)$, and is independent of initiation rate $\alpha$ and termination rate $\beta$, see Fig. \ref{Fig.Analysis_vs_Simulation_K_is_not_1_ex}{\bf (f)}. Properties of single species density $\rho_k$ may be different from $\rho$. For example, if $\rho$ lies in $[0, 1/2]$ or $[\rho_l, 1]$, it will increase along the track. Otherwise, $\rho$ decreases along the track.
This is because that $\partial\rho=(K+1)\Omega(\rho-\rho_l)/(2\rho-1)$, see Eq. (S20) in \cite{supplemental}. Therefore, if there exists DW in $(0,1)$ and the termination rate $1-\rho_l<\beta<1$, then after DW location $x_w$, density $\rho$ decreases monotonically. Otherwise, if $0<\beta<1-\rho_l$, $\rho$ will increase after $x_w$. However, the results in Fig. \ref{Fig.Analysis_vs_Simulation_K_is_not_1_ex}{\bf (b,c)} show that, after DW location $x_w$, the monotonicity of $\rho_k$ may be different from $\rho$.
\begin{figure}[!htp]
  \centering
  \includegraphics[width=8.6cm]{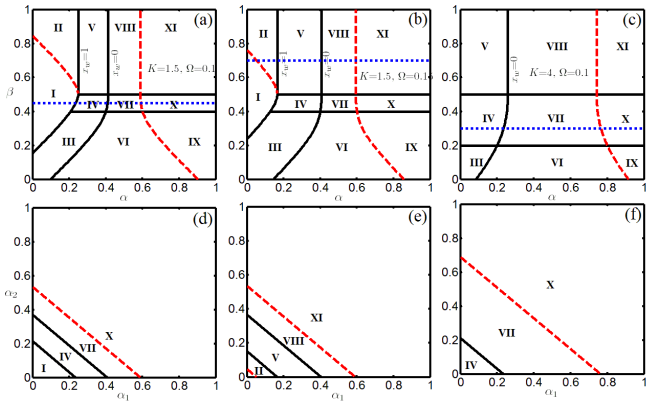}\\
  \caption{(Color online) Phase diagrams of total density $\rho$ obtained by  stationary mean-field Eq.(\ref{EqBulk00:Sum}) for $K>1$. Parameter values used in calculations are
  $\Omega=0.1$, $K=1.5$ for {\bf (a,d)};
  $\Omega=0.15$, $K=1.5$ for {\bf (b,e)}; and
  $\Omega=0.1$, $K=4$ for {\bf (c,f)}.
  {\bf (d,e,f)} are phase diagrams in $(\alpha_1, \alpha_2)$-plane with termination rate $\beta=0.45,0.7,0.3$, i.e., corresponding to horizontal dotted lines in {\bf (a,b,c)}, respectively. The meanings of Roman numerals in each connected area of figures are as follows.
  (1) {\bf I}:=LD-BL$_r^+$, which means that density $\rho$ is in LD phase, and there exists {\it right} boundary layer (BL) in which $\rho$ {\it increases} sharply, i.e. $\rho(1)>\rho(1-\epsilon)$ for small real number $\epsilon$. (2) {\bf II}:=LD-BL$_r^-$, i.e., $\rho$ is in LD phase, and it {\it decreases} sharply at right boundary.
  (3) {\bf III}:=LD-DW-HD$_1$, near left boundary density $\rho$ is in low density phase, while near right boundary $\rho$ is in high density phase, and there exists DW between these two phases. Here \lq\lq HD$_1$'' means  $\rho>\rho_l$, i.e., $\beta<1-\rho_l$.
  (4) {\bf IV}:=LD-DW-HD$_2$, where HD$_2$ means that density $\rho$ is between $0.5$ and $\rho_l$, i.e., $1-\rho_l<\beta<0.5$.
  (5) {\bf V}:=LD-DW-M$_r$, where M$_r$ means that density $\rho$ is between $0.5$ and $\rho_l$ but its value is independent of right boundary condition, therefore right BL appears. M$_r$ can be regarded as the right half part of \lq\lq Meissner'' phase. For this phase, $\beta>0.5$.
  (6) {\bf VI}:=BL$_l^+$-HD$_1$, i.e., $\rho>\rho_l$ and $\rho$ increases in left BL.
  (7) {\bf VII}:=BL$_l^+$-HD$_2$. (8) {\bf VIII}:=M$^+$, which means that $\rho$ is in \lq\lq Meissner" phase and it increases in left BL.
  (9) {\bf IX}:=BL$_l^-$-HD$_1$. (10) {\bf X}:=BL$_l^-$-HD$_2$. (11) {\bf XI}:=BL$_l^-$-M. VIII and XI are two species cases of \lq\lq Meissner" phase. For examples of density $\rho$ which is in one of the above eleven phases, see Fig. S4 in \cite{supplemental}. 
  }
  \label{Fig.Kisnot1Phase}
\end{figure}

Previous analysis about special cases $K=1$ has shown that, if total density $\rho$ is in MC phase, i.e. $\rho\equiv1/2$, then single species density $\rho_k$ may not be constant. Similar results hold for the general $K>1$ cases. If $\beta=1-\rho_l$, then $\rho\equiv\rho_l=K/(K+1)$ is constant near right boundary $x=1$. The plots in Fig. S5 show that the corresponding single species densities $\rho_k$ may not be constant. Theoretical analysis gives that the monotonicity of $\rho_k$, when $\rho\equiv\rho_l$ is constant, is also determined by the sign of $\Omega_{1,A}\alpha_2-\Omega_{2,A}\alpha_1$. With positive values of it, $\rho_1$ increases, while $\rho_2$ decreases, along motion track. Both densities $\rho_1$ and $\rho_2$ will be constant iff $\Omega_{1,A}\alpha_2-\Omega_{2,A}\alpha_1=0$, see \cite{supplemental} for detailed analysis.

Examples of phase diagram of total density $\rho$ in $(\alpha, \beta)$ plane, for general cases $K>1$, are plotted in Fig. \ref{Fig.Kisnot1Phase}{\bf (a,b,c)}. Similar as in the one-species \lq\lq TASEP-LK", BL may appear at one or both of the two boundaries. DW may appear in interval $(0,1)$, and density $\rho$ may be in LD phase ($\rho<1/2$) or HD phase ($\rho>1/2$). To show more details about the \lq\lq TASEP-LK" process, in Fig. \ref{Fig.Kisnot1Phase} the HD phase is divided into two different cases, HD$_1$ phase ($1/2<\rho<\rho_l=K/(k+1)$) and HD$_2$ phase ($\rho>\rho_l$). From the phase diagram in $(\alpha, \beta)$ plane, phase diagrams in any planes of parameter pair $(\sigma_1, \sigma_2)$  can be easily obtained. Where $\sigma_k=\alpha_1, \alpha_2, \beta_1, \beta_2$. Examples of phase diagram in $(\alpha_1,\alpha_2)$ plane are plotted in Fig. \ref{Fig.Kisnot1Phase}{\bf (d,e,f)}, which are corresponding to the dotted horizontal lines in Fig. \ref{Fig.Kisnot1Phase}{\bf (a,b,c)} respectively. The above discussion about the relationship between total density $\rho$ and single species density $\rho_k$ implies that phase diagrams of density $\rho_k$ are the same as those of the total density $\rho$.

In summary, \lq\lq TASEP-LK" process with two particle species is discussed in this study. Different from previous studies about two-species TASEP, particle attachment/detachment to/from bulk sites of motion track is allowed, both of the two particle species enter into the track from the same boundary, and move unidirectionally to the same direction. The two particle species do not change to each other, and do not change their locations even if they are adjacent to each other \cite{Evans1995,Mobilia2001,Schutz2003,Zeraati2013}. This study found that, domain walls and boundary layers of total particle density and single species densities always appear simultaneously. The height of domain wall of total density is equal to the summation of those of the two single species. Based on these properties, theoretical methods to obtain steady state densities of the two particle species are presented. Our results show that properties of single species densities may be different from that of the total particle density. In this study, phase diagrams of particle density in typical parameter planes are also presented. The methods presented in this study are also available to the analysis of \lq\lq TASEP-LK" process including more than two particle species. The results of this study are helpful to further understandings of the biophysical process of cargo transportation in living cells, where one single protofilament of microtubule is actually occupied by various kinds of cargos and motor proteins \cite{Howard2001,Vale2003}.

\begin{acknowledgments}
This study was supported by the Natural Science Foundation of China (Grant No. 11271083), and the National Basic Research Program of China (National \lq\lq973" program, project No. 2011CBA00804).
\end{acknowledgments}


\end{document}